\documentclass[12pt]{article}
\usepackage[utf8]{inputenc}
\usepackage[T1]{fontenc}
\usepackage[english]{babel}
\usepackage{graphicx}
\usepackage{amsmath}

\title{Study of simulated Bloch oscillations  in strained graphene using neural networks.}

\author{\textbf{J.A. Gonz\'alez, C. E. L\'opez, A. Raya}\\
Laboratorio de Inteligencia Artificial y Superc\'omputo\\
Instituto de F\'isica y Matem\'aticas \\
Universidad Michoacana de San Nicol\'as de Hidalgo \\
Ciudad Universitaria. Edificio C-3\\
C.P. 58040\\
Morelia, Michoac\'an\\}
\date{August 2018}

\begin{document}

\maketitle
\begin{abstract}
We consider a monolayer of graphene under uniaxial, tensile strain and simulate Bloch oscillations for different electric field orientations parallel to the plane of the monolayer using several values of the components of the uniform strain tensor, but keeping the Poisson ratio in the range of observable values. 
We analyze the trajectories of the charge carriers with different initial conditions using an artificial neural network, trained to classify the simulated signals according to the strain applied to the membrane. When the electric field is oriented either along the Zig-Zag or the Armchair edges, our approach successfully classifies the independent component of the uniform strain tensor with up to 90\% of accuracy and an error of $\pm1\%$ in the predicted value. For an arbitrary orientation of the field, the classification is made over the strain tensor component and the Poisson ratio simultaneously, obtaining up to 97\% of accuracy with an error that goes from $\pm5\%$ to $\pm10\%$ in the strain tensor component and an error from $\pm12.5\%$ to $\pm25\%$ in the Poisson ratio.\\

\end{abstract}

\section{Introduction}

Modern material science has received a tremendous impact after the first isolation of graphene membranes~\cite{novo1}, giving rise to the era of 2D materials. Graphene possesses a number of outstanding properties, ranging from tremendously high electric and thermal conductivities, transparency of the membranes and, on top of that, stiffness and flexibility~\cite{rise1,rise2,rise3,rise4}. Thus, the manipulation of electric properties through mechanical means has given rise to the field of straintronics~\cite{straintronics} in graphene and other materials (see~\cite{review} for a recent review). On theoretical grounds, mechanical deformations of graphene membranes are usually accounted for through a strain tensor that describes the deviation of the graphene curvature with respect to the ideal flat case. The effect is then seen in tilting and displacing of the Dirac points in reciprocal space plus a re-shaping of these points such that the isoenergetic contours of these cones is elliptical, namely, the Fermi velocity becomes anisotropic and of tensor nature~\cite{review}. In the limiting case of uniform, tensile strain, these features might be completely understood as a strain modified reciprocal lattice such that the dispersion relation is modified from the pristine case by different constants (related to the components of the strain tensor) along the Armchair and Zig-Zag directions that in the low energy limit account for the anisotropy of the Fermi velocity~\cite{maurice,yajaira}. Considering this setup, in this article we explore the impact of strain in the scenario of Bloch oscillations (BO) in monolayer graphene. 

BO are a remarkable phenomenon in traditional solid state physics. In spite of the fact that these oscillations are not directly observed in real solids, its study demonstrates the influence of a periodic array in conjunction with an external force field in the motion of charge carriers in different materials. The observation of BO has been done under several experimental settings in high-purity semiconductor superlattices~\cite{slt1,slt2,slt3,slt4,slt5,slt6,slt7,slt8,slt9,slt10}, atomic systems~\cite{as1,as2}, dielectric~\cite{die1,die2,die3}, plasmonic waveguide arrays~\cite{plas} and also in  bilayer graphene superlattices~\cite{bil1,bil2}. All these observations make it relevant to study this phenomenon beyond solids.  
In this connection the inverse problem of BO has already been addressed by our group for the linear chain~\cite{chain}, the 2D square lattice~\cite{square} and pristine graphene~\cite{graph} through an Artificial Neural Networks (ANN) approach. In this article we extend and generalize these findings to the case of graphene under uniaxial strain. As compared to the pristine case, the first natural difference that appears under strain is the change of the period of oscillations. Moreover, it is observed that the amplitudes of closed trajectories change in such a way that new self-intersecting patterns appear~\cite{saul}. Considering that machine learning methods are a consistent and reliable source to identify and classify patterns in general, we believe that a pointwise study of the modification of these oscillations by mechanical deformations of the membrane and its complete characterization using a specific setup of ANNs, is a natural question to be addressed.

ANN is part of the called machine learning methods, that actually are present in our daily life: the smartphones have facial, voice and fingerprint recognition, suggestion systems for the music and movies that we like, weather prediction, autonomous driving of vehicles and in many other applications. These machine learning methods learn to respond depending on the supplied data, where this data could represent information of practically any problem. For this reason they are used in several areas of science, in particular we have used the methods to analyze different physical systems, for example, gamma ray burst ~\cite{GRBs}, obstruction detection inside pipes ~\cite{pipes} and gravitational waves ~\cite{GWs1,GWs2} to mention some topics. 

In this work we address the inverse problem of BO in uniaxially strained graphene under uniform extension of the membrane. We start in Section~\ref{sec:strain} by describing the dispersion relation of graphene under strain and the issue of BO at the semiclassical level. We further specify our considerations for the simulation of these BO in Section~\ref{sec:sim}. The structure of the ANN is discussed in Section~\ref{sec:ANN} and results presented in Section~\ref{sec:res}. We finally conclude in Section~\ref{sec:conclu}.

\section{Bloch oscillations in strained graphene}\label{sec:strain}

\begin{figure}[h]
	\centering
	\includegraphics[width=0.45\columnwidth]{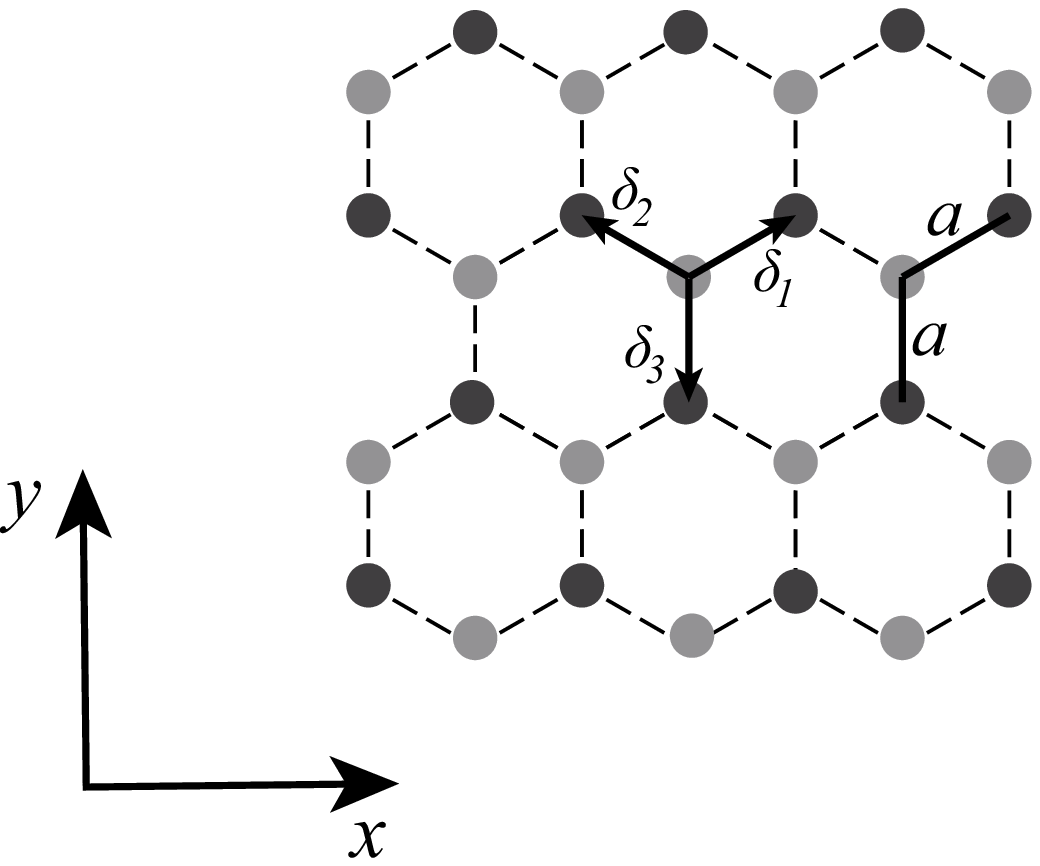}
  \includegraphics[width=0.45\columnwidth]{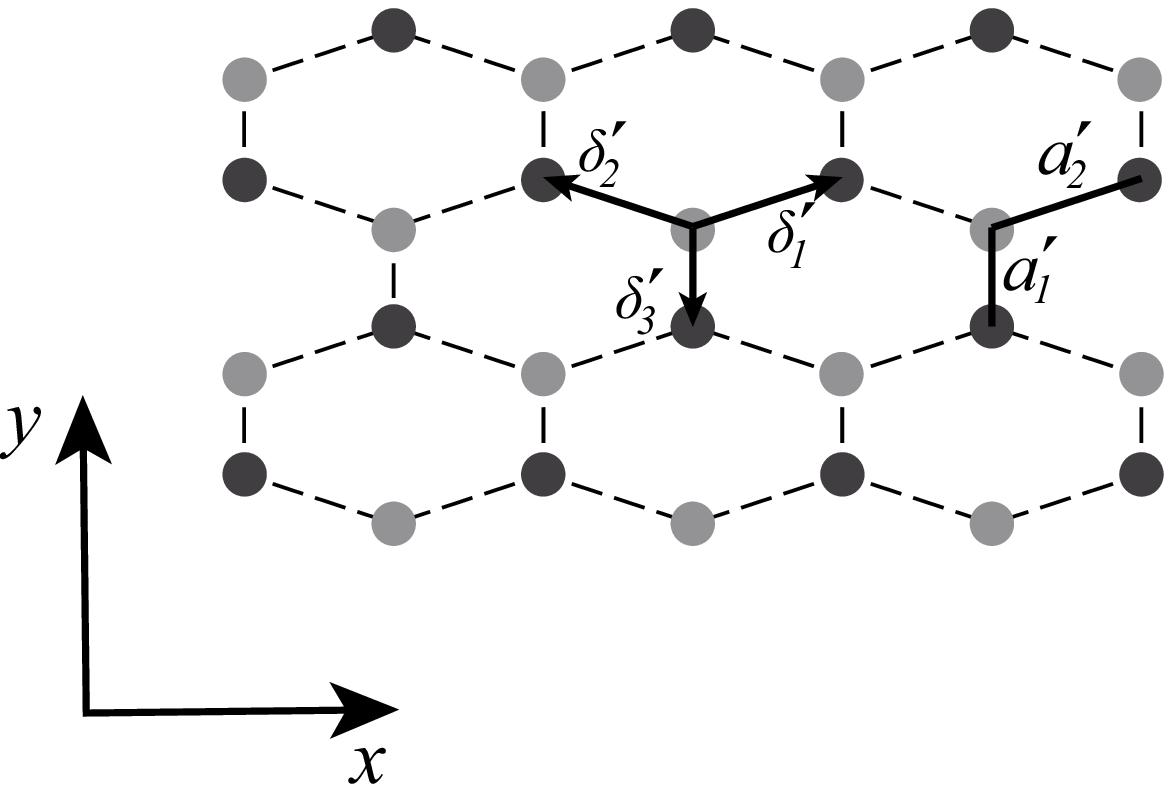}
	\caption{Crystal structure of graphene. {\em Left panel:} Pristine case. {\em Right panel:} Under uniaxial, tensile strain.}
	\label{crystal}
\end{figure}

We consider the connection between elasticity and the tight-binding description of graphene following closely reference~\cite{maurice}. Considering the honeycomb array of graphene as the superposition of two triangular sublattices, the position of the atoms in the deformed sublattice $A$ (see Fig.~\ref{crystal}) can be written as $\mathbf{x}'=(I+\epsilon)\mathbf{x}$, where $I$ is the identity matrix and  $\epsilon$ is the coordinate independent strain tensor. Thus, the nearest-neighbor tight-binding Hamiltonian is expressed as
\begin{equation}
H=-\sum_{\mathbf{x}',n} t_{n} a^\dagger_{\mathbf{x}'} b_{{\mathbf{x}'+\boldsymbol{\delta}_n'}}+h.c.,\label{tbs1}
\end{equation}  
where $\mathbf{x}'$ runs over all the grid points of the sublattice $A$ and $\boldsymbol{\delta}_n'$ are the vectors connecting every point with the nearest neighbors. Here, $a^\dagger_{\mathbf{x}'}$ and $b_{{\mathbf{x}'+\boldsymbol{\delta}_n'}}$ are the creation and anihilation operators for charge carriers in the sublattices $A$ and $B$ respectively at the corresponding sites ${\mathbf{x}'}$ and ${\mathbf{x}'+\boldsymbol{\delta}_n'}$. Notice that the hopping parameters $t_n$ in the Hamiltonian~(\ref{tbs1}) are considered coordinate independent, assumption that is valid only for the case of uniform strain. Upon Fourier expanding  the creation and anihilation  operators, in momentum space, the tight-binding Hamiltonian has a form very similar to the ideal case, namely,
\begin{eqnarray}
H=-\sum_{\mathbf{k},n} t_n e^{-i\mathbf{k}\cdot(I+\epsilon)\cdot \delta_n}a_{\mathbf{k}}^\dagger b_{\mathbf{h}}+ h.c.,
\end{eqnarray}
with the difference that the hopping parameters are now position-dependent. The dispersion relation is straightforwardly obtained as
\begin{equation}
\varepsilon(\mathbf{k})= \pm \left|\sum_n t_n e^{-i\mathbf{k}^*\cdot \delta_n} \right|,
\end{equation}
with $\mathbf{k}^*=(I+\epsilon)\cdot\mathbf{k}$. At the linear order on the strain tensor, we write
\begin{equation}
t_n=t_0\left(1-\frac{\beta}{a^2}\boldsymbol{\delta}_n\cdot \epsilon \cdot  \right)\boldsymbol{\delta}_n,
\end{equation}
where $t_0$ is the hopping parameter of pristine graphene and $\beta\simeq3$ is the variation of the hopping energy due to lattice deformation. Then, using that
\begin{equation}
\boldsymbol{\delta}_1= \frac{a}{2}(\sqrt{3},1),\quad \boldsymbol{\delta}_2=\frac{a}{2}(-\sqrt{3},1),\quad \boldsymbol{\delta}_3=a(0,-1),
\end{equation}
with $a$ the interatomic distance in an ideal sample, we explicitly have
\begin{equation}
\varepsilon(\mathbf{k})=\pm t_0 \sqrt{3+f(\mathbf{k}^*) -\beta(3{\rm Tr}(\epsilon)+f_\epsilon(\mathbf{k}^*)+\beta^2f_{\epsilon^2}(\mathbf{k}^*))},\label{fulldr}
\end{equation}
where
\begin{equation}
f(\mathbf{k}^*)=2\cos\left(\sqrt{3}k_x^* a \right) + 4 \cos\left( \frac{\sqrt{3}k_x^*a}{2}\right)\cos\left( \frac{3k_y^*a}{2}\right),
\end{equation}
whereas $f_\epsilon(\mathbf{k}^*)$ and $f_{\epsilon^2}(\mathbf{k}^*)$ represent modifications of the spectrum at first and second order in $\beta$, respectively. For the analysis in this paper,  we consider $\beta=0$ and use the simplified dispersion relation
\begin{equation}
\varepsilon(\mathbf{k})=\pm t_0 \sqrt{3+f(\mathbf{k}^*)}\label{dispr}
\end{equation} 
and leave the full dispersion relation for a future work.

For the analysis of BO, we consider the semiclassical equation of motion
\begin{equation}
\frac{d\mathbf{k}}{dt}=-e\mathbf{E},
\end{equation}
where $\mathbf{E}$ represents a static, uniform electric field and $e$ is the fundamental charge. After integration, we obtain the expresion $\mathbf{k}(t)=\mathbf{k}(0)-e\mathbf{E} t$ and we substitute it into the dispersion relation~(\ref{dispr})
\begin{equation}
\frac{d\mathbf{r}}{d t}= \frac{\partial \varepsilon(\mathbf{k})}{\partial \mathbf{k}} \, .
\label{eq:pos}
\end{equation}
Integrating this equation we can obtain the position of the charge carriers at a given time $t$. We consider a strain tensor of the form
\begin{equation}
\epsilon=\left( 
\begin{array}{cc} \epsilon_{xx} & 0 \\  0 & -\nu \epsilon_{xx} \end{array}
\right)\;,
\end{equation}
with the Poisson ratio $\nu$. Below we detail the procedure to simulate BO in strained graphene from this framework.

\section{Simulated Bloch oscillations}\label{sec:sim}
Once we have the equations that describe the position of the electric charge carriers as a function of time, we need to specify some initial conditions like the initial momentum ($k_x(0)$, $k_y(0)$) or the external electric field ($E_x$, $E_y$) and simulate the trajectories of the carriers for a fixed lapse, only varying the strain. Additionally we consider that $\hbar=a=\tau=1$ and $e=-1$ for a time interval of $T=4 \pi$ units. Notice that with these assumptions, the analyzed quantities do not have physical units.\\
\\
Two cases are studied according to the parameters that are varied when the oscillations are generated numerically:
\begin{enumerate}
    \item The only parameter that varies is $\epsilon_{xx}$ with $\epsilon_{xy}=\epsilon_{yx}=0$, $\nu=0.16$ and the other parameters fixed in three subcases
    \begin{itemize}
        \item[a)] $E_x=1,$ $E_y=0,$ $k_x(0)=0, $ $k_y(0)=\pi/\sqrt{3}$.
        \item[b)] $E_x=0,$ $E_y=1,$ $k_x(0)=\pi/\sqrt{3},$ $k_y(0)=0$.
        \item[c)] $E_x\ne 0$, $E_y\ne 0$, $k_x(0)=k_y(0)=0$.
    \end{itemize}
    
    \item The parameters that vary are $\epsilon_{xx}$ and $\nu$ with $E_x\ne 0,$ $E_y\ne 0,$ $k_x(0)=k_y(0)=0$ and $\epsilon_{xy}=\epsilon_{yx}=0$.
\end{enumerate}

In the first case, $N$ different values are used for $\epsilon_{xx}$ and they are equidistant in the interval [0, 0.25] and labeled in $C_{\epsilon}$ classes or groups, such that each class has the same number of simulations, {\it i.e.} $\mod(N/C_{\epsilon})=0$ must be satisfied. The ANN will perform a classification using as input data for training, the components $x(t)$ and $y(t)$ of the electric carrier position. We interpret each predicted class as a value for $\epsilon_{xx}$ with a relative error related with the total length of the interval $\epsilon_L=0.25$. We define
\begin{equation}
    \hat{\epsilon}_{m}=(2m-1)\epsilon_L/2C_{\epsilon} \pm \epsilon_L/2C_{\epsilon},\hspace{.5cm} 1\le m \le C_{\epsilon},
    \label{classes1}
\end{equation}
as  the predicted value of $\epsilon_{xx}$ associated with the class $m$, where $C_{\epsilon}$ is the total number of classes.\\

The second case is similar to the first one, but now also the parameter $\nu$ is varied in the interval [0, 0.2] selecting $N'$ different equidistant values and grouping them in $C_{\nu}$ classes. In this scenario the total number of generated patterns is $N\times N'$ and the ANN classifies both of the parameters simultaneously: $\epsilon_{xx}$ is associated to one class from the total of $C_{\epsilon}$ and $\nu$ to one class from $C_{\nu}$ classes. The predicted value for $\nu$ has a similar expression as in Eq. (\ref{classes1})
\begin{equation}
    \hat{\nu}_{n}=(2n-1)\nu_L/2C_{\nu} \pm \nu_L/2C_{\nu},\hspace{.5cm} 1\le n \le C_{\nu}, 
    \label{classes2}
\end{equation}
with $\nu_L=0.2$ the total length of the interval where $\nu$ is varied.\\

In Eqs. (\ref{classes1}) and (\ref{classes2}) we observe that as we increase the number of classes, the error associated with each prediction is smaller. It is worth to mention that as the errors in the predictions decrease, also the  efficiency in the classification decreases, as we will illustrated in the next section. 
This numerical approach, where simulations are generated and classifications are studied considering different values for the parameters and the number of classes,was previously used in \cite{chain,square,graph} where are studied Bloch oscillations in simpler physical systems.\\

\section{Artificial neural networks}\label{sec:ANN}
We use a feedforward ANN to classify patterns such that the network estimates the parameters $\epsilon$ and $\nu$ that generated the numerical simulations. To train the ANN first it is required to preprocess the data that will feed into the network. Also, as we are working with a supervised learning algorithm, we have to define the targets associated with each one of the patterns.\\
The input data used, is obtained from two time series: the position $x(t)$ and $y(t)$ of the charge in a simulated BO in the time interval $0\le t \le T$ (where $T$ is the total duration of the oscillation), subject to different imposed conditions and the position is obtained integrating numerically the Eq. (\ref{eq:pos}). We have divided the simulation in fifty steps in time such that $t_i=i\Delta t$, with $0\le i \le 49$ and $\Delta t=T/50$. 
The input vector for each pattern $p$ is constructed as
\begin{equation}
I^p=\{x(t_0),y(t_0),\ldots,x(t_{49}),y(t_{49})\}, \hspace{.1cm}1 \le p \le N_p, 
\label{inputs}
\end{equation}
with $N_p$ the total number of patterns.  As mention in the previous section, $N_p=N$ in the first case and $N_p=N\times N'$ in the second. From the total number of patterns, seventy percent of them are chosen randomly to train the network and the remaining thirty percent corresponds to the validation set. The purpose of this validation set is to avoid that the training process reaches a state of overtraining, producing an excellent behavior during  the prediction of the training set, but having a bad performance over patterns not used in the training. To test the performance of the network, the same number of signals as in the validation set are simulated but this time using random values of the variables $\epsilon_{xx}$ and $\nu$, ensuring that the new simulations are inside the range under consideration. The selection of the training and validation sets is made once that the target values for the patterns have been prepared. \\
The activation functions used in the hidden and output layers are sigmoid functions. These are chosen such that the output values are in the open interval (0,1). Then, we define the target values for $C_{\epsilon}$ and $C_{\nu}$, representing the different classes in the range of the function. The proposed targets associated with each pattern and class are:
\begin{eqnarray}
    \hat{T}_{\epsilon_m}=(2m-1)/(2C_{\epsilon}),\hspace{.3cm} 1\le m \le C_{\epsilon},\nonumber\\
    \hat{T}_{\nu_n}=(2n-1)/(2C_{\nu}),\hspace{.3cm} 1\le n \le C_{\nu},
\end{eqnarray}
where $\hat{T}_{\epsilon_m}$ is the target for all the patterns that are in the class $m$ and $\hat{T}_{\nu_n}$ is the target for all the patterns that are in the class $n$. For example, if the ANN is fed with one pattern among the first $N/C_{\epsilon}$ it will be part of the class $m=1$ and the corresponding target is $\hat{T}_{\epsilon_1}=1/2C_{\epsilon}$.\\

The ANN was programmed in FORTRAN 90 and trained with an offline supervised backpropagation learning algorithm designed to minimize a cost function type mean squared error \cite{rojas,bishop}.  The generation of the simulated BOs, the preprocessing of the data and also the visualization of the results was performed with Wolfram Mathematica.\\
The structure of the network has one input layer with two hundred neurons that receive the extracted data from each pattern, one hidden layer with a variable number of neurons to be determined according to ANN performance and one output layer with one neuron in the first case, and two neurons in the second case. One of the output neurons produce a value that is related to $\epsilon_{xx}$ and the other to $\nu$. For this reason the structure of the output layer changes depending on the case.\\

With the input data, the desired outputs and the network structure ready, it is necessary to train the network for a suitable number of iterations and evaluate the performance of ANN counting the number of predictions correctly classified. We consider that the pattern has been correctly classified if the output of the network corresponding to this pattern has an output value between $\hat{T}_{\epsilon_m}-1/2C_{\epsilon}$ and  $\hat{T}_{\epsilon_m}+1/2C_{\epsilon}$. A similar condition is employed to define a correct classification when the output corresponding to the variable $\nu$ is in the class $n$: if the output is between $\hat{T}_{\nu_n}-1/2C_{\nu}$ and  $\hat{T}_{\nu_n}+1/2C_{\nu}$ the pattern is considered to be correctly classified.\\

In the next section we present the results obtained after the ANNs have been trained, considering variations in some of the parameters of the networks, for instance,  the number of hidden neurons, the learning rate and/or the number of classes in which were divided the simulations.

\section{Results}\label{sec:res}
The results presented in this section are those generated by the network that has obtained the lower cost at the end of the training from all the network parameters considered, where the number of hidden neurons was equal to 2$^j$ for $2\le j\le 6$ and the learning rate was equal to 3$^{-l}$ for $3\le l \le 7$, with $j$ and $l$ integers. This lead to a total of 25 combinations of the parameters explored with a parallelized code using MPI.\\
A total of $2\times 10^4$ learning iterations were used, the number of simulations in each case is of 10$^3$, and (depending on the case) the number of classes used is different.
\subsection{Prediction of $\epsilon_{xx}$}
\begin{figure}
	\centering
	\includegraphics[width=8.5cm]{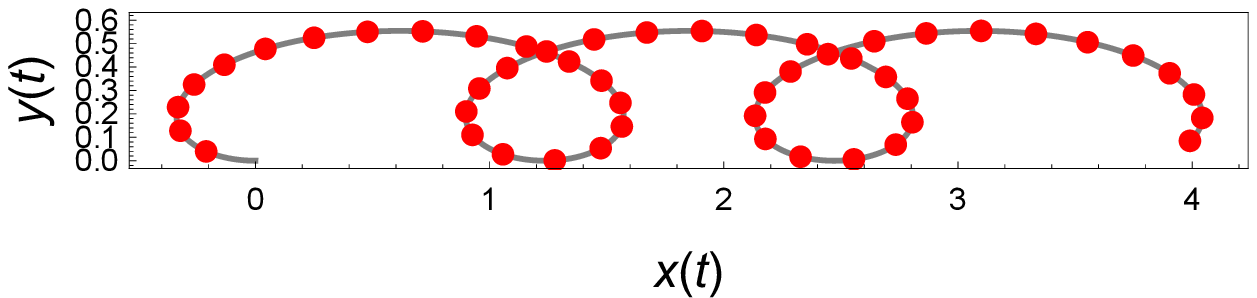}
	\caption{Sample of the trajectory of an electron when $\epsilon_{xx}=0.20$ during the time interval $0\le t \le 4\pi$ and the initial conditions considered for the case 1a. The red dots represent the coordinates used as inputs for the ANN.}
	\label{trajectorycase1a}
\end{figure}

\begin{figure}
	\centering
	\includegraphics[width=8.5cm]{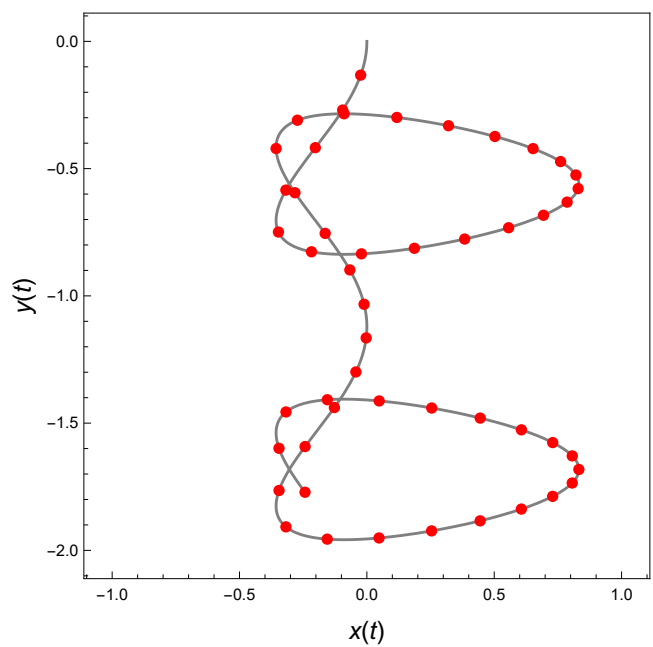}
	\caption{Sample of the trajectory of an electron when $\epsilon_{xx}=0.20$ during the time interval $0\le t \le 4\pi$ and the initial conditions considered for the case 1b. The red dots represent the coordinates used as inputs for the ANN.}
	\label{trajectorycase1b}
\end{figure}

\begin{figure}
	\centering
	\includegraphics[width=8.5cm]{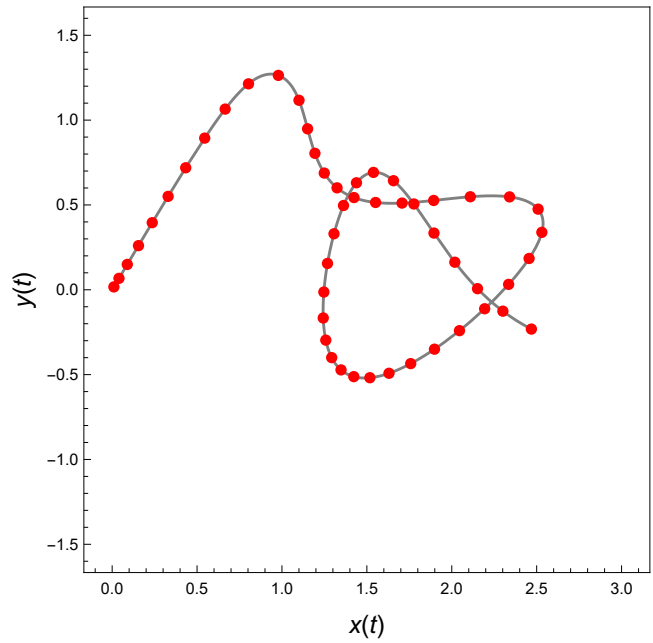}
	\caption{Sample of the trajectory of an electron when $E_x=-0.33$, $E_y=-0.79$ and $\epsilon_{xx}=0.20$ during the time interval $0\le t \le 4\pi$ and the initial conditions considered for the case 1c. The red dots represent the coordinates used as inputs for the ANN.}
	\label{trajectorycase1c}
\end{figure}

The first case includes simulations of the BOs with the initial conditions mentioned in Section~\ref{sec:sim} and were classified in $C_{\epsilon}=50, 100$ and $200$ classes.\\
A sample of the data chosen as input for the network according to the Eq. (\ref{inputs}) for the subcases a), b) and c) are in the Figs. \ref{trajectorycase1a}, \ref{trajectorycase1b} and \ref{trajectorycase1c} respectively.\\

Once we have selected the network with the parameters that produces the lowest cost during training, the network performance is evaluated calculating the Percentage of Predictions Correctly Classified (PPCC) in each case and the results are displayed
in Table \ref{results1}. We can observe that for each of the subcases, the more considered classes, the lower the PPCC as it is expected. We also observe, in one hand that the best performance is obtained in case 1a), and in the other the case 1c) (where both components of the electric field are different from zero) it has the lower PPCC. As mentioned before, although the PPCC is lower when the number of classes increases, the error associated to the predicted average value is lower as established Eq. (\ref{classes1}). Hence the error associated with the average value $\hat{\epsilon}$ is of $\pm 1\%$ of $\epsilon_L$ when the simulations are divided in fifty classes, meanwhile the error associated is of $\pm 0.25\%$ of $\epsilon_L$ when two hundred classes are selected. Depending on the accuracy needed, we can chose the number of classes in which the patterns are divided.

\begin{table}[h]
	\begin{center}
		\begin{tabular}{|c|c|c|c|}
			\hline
			\multicolumn{4}{|c|}{Case 1a}\\
			\hline 
			PPCC(\%) for &\hspace{.1cm}Training\hspace{.1cm}&Validation&\hspace{.3cm}Test\hspace{.3cm}  
			\tabularnewline
			\hline
		    $C_{\epsilon}=50$& 90.1 &90.0 &88.6 \tabularnewline
			$C_{\epsilon}=100$& 82.5& 77.6 &86.0 \\
			$C_{\epsilon}=200$ & 82.1  & 77.0 &79.0\tabularnewline
			\hline
			\hline
			\multicolumn{4}{|c|}{Case 1b}\\
			\hline 
			$C_{\epsilon}=50$& 88.1 & 86.6& 90.3\tabularnewline
			$C_{\epsilon}=100$& 80.2&75.6  &79.6 \\
			$C_{\epsilon}=200$ & 68.4 & 65.0 &68.3\tabularnewline
			\hline
			\hline
			\multicolumn{4}{|c|}{Case 1c}\\
			\hline
		    $C_{\epsilon}=50$& 81.8 & 86.6&81.6 \tabularnewline
			$C_{\epsilon}=100$& 66.7& 65.3 & 69.0\\
			$C_{\epsilon}=200$ &  59.5 & 52.6 & 54.6 \tabularnewline
			\hline
		
		\end{tabular}
		
		\caption{PPCC obtained by the ANN for the training, validation and test sets in case 1. The output generated by the network predicts the value of $\epsilon_{xx}$.}
		\label{results1}
	\end{center}
\end{table}

\subsection{Prediction of $\epsilon_{xx}$ and $\nu$}
\begin{figure}
	\centering
	\includegraphics[width=8.5cm]{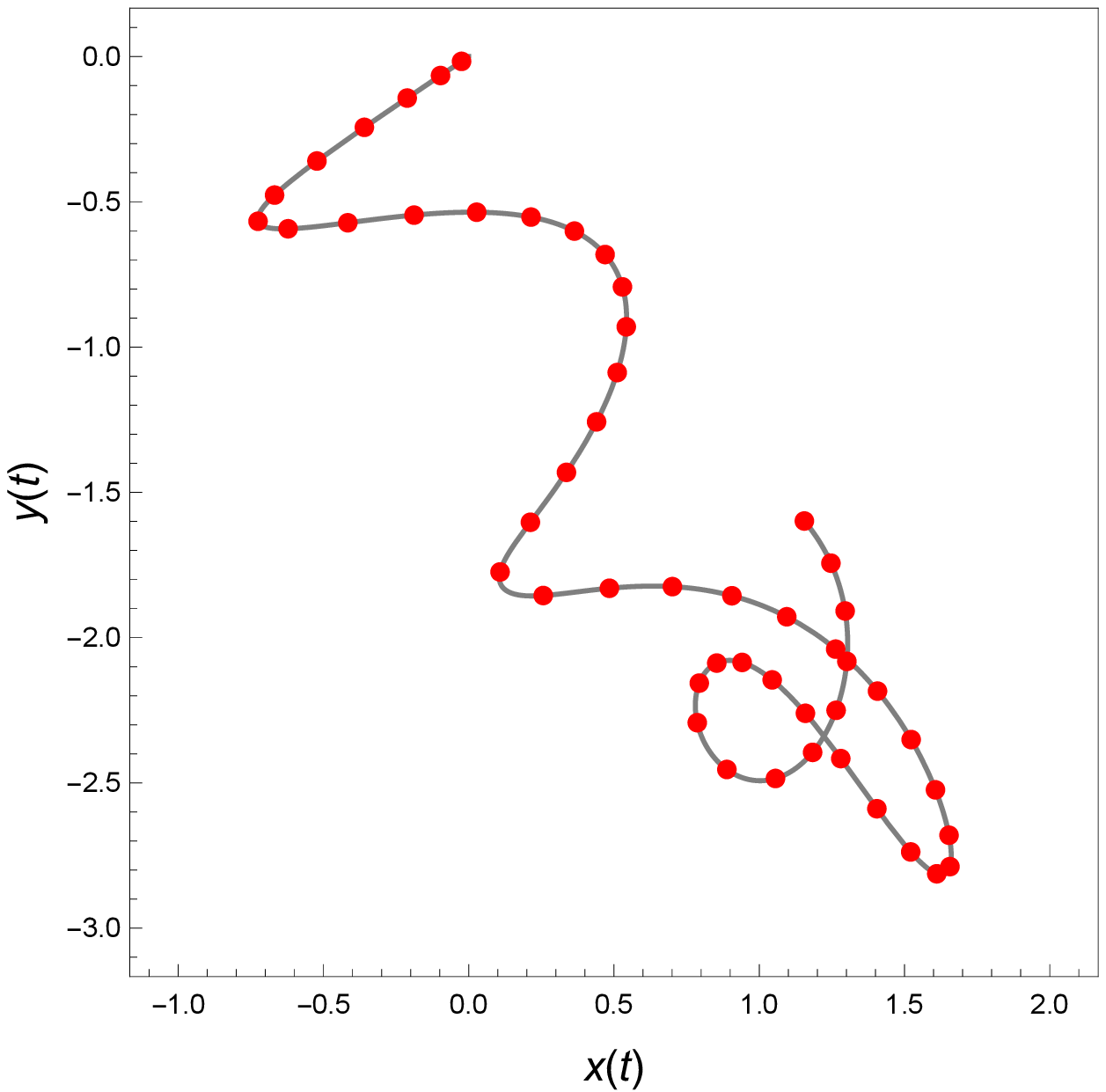}
	\caption{Sample of the trajectory of an electron when $E_x=0.76$, $E_y=0.78$, $\epsilon_{xx}=0.20$, and $\nu=0.16$ during the time interval $0\le t \le 4\pi$ and the initial conditions considered for the case two. The red dots represent the coordinates used as inputs for the ANN.}
	\label{trajectorycase2}
\end{figure}

For the second case of study we created the same amount of simulations as in the previous case but selecting $N=50$ and $N'=20$, varying $\epsilon_{xx}$ and $\nu$ respectively and considering $C_{\epsilon}=5$ and $10$ classes for $\epsilon_{xx}$ and $C_{\nu}=2$ and $4$ classes for $\nu$. $E_x$ and $E_y$ were selected randomly in the interval [-1,1], showing a sample of the trajectory for the considered lapse of time and initial conditions in Fig. (\ref{trajectorycase2}). 

The PPCC for this case are shown in the Table \ref{results2}.
\begin{table}[h]
	\begin{center}
		\begin{tabular}{|c|c|c|c|c|}
			\hline
			\multicolumn{5}{|c|}{Case 2}\\
			\hline
			\hline 
			PPCC(\%) for &Output&\hspace{.1cm}Training\hspace{.1cm}&Validation&\hspace{.3cm}Test\hspace{.3cm}  
			\tabularnewline
			\hline
		    $C_{\epsilon}=5$& ${O}_{1}$ & 99.8 & 100.0 & 96.3\tabularnewline
			$C_{\nu}=2$ & $O_2$ &99.0 & 98.0 &97.3 \\
			\hline
			\hline
			$C_{\epsilon}=5$ & ${O}_{1}$ &  99.4  & 98.0 &97.6\tabularnewline
			$C_{\nu}=4$ & $O_2$&91.4 & 91.3 &88.3 \\
			\hline
			\hline
			$C_{\epsilon}=10$ & ${O}_{1}$ & 95.1 & 96.0&92.6\tabularnewline
			$C_{\nu}=2$ & $O_2$ &92.7& 92.1& 94.0\\
			\hline
			\hline
			$C_{\epsilon}=10$ & ${O}_{1}$ & 98.5&98.0  & 84.3\tabularnewline
			$C_{\nu}=4$ & $O_2$ & 83.1& 83.6&87.0 \\
			\hline
		
		\end{tabular}
		
		\caption{PPCC obtained by the ANN for the training, validation and test sets in  case 2. The output $O_1$ predicts the value of $\epsilon_{xx}$ and the output $O_2$ the one of $\nu$. Both predictions were divided in $C_{\epsilon}$ and $C_{\nu}$ classes respectively.}
		\label{results2}
	\end{center}
\end{table}

The behavior of the PPCC is similar to case 1, obtaining more correctly classified patterns when less classes are considered. As the number of simulations is the same as in the previous case, but with two parameters that vary, the total number of classes for each parameter is less. As a consequence, the associated errors for the predicted value of $\epsilon_{xx}$ are $\pm5\%$ and $\pm10\%$ of $\epsilon_{L}$ for 10 and 5 classes respectively and an error for the predicted value of $\nu$ of $\pm12.5\%$ and $\pm25\%$  of $\nu_{L}$  for 4 and 2 classes respectively.

\section{Conclusions}\label{sec:conclu}

In this work we have considered the inverse problem of BO in strained graphene. We have considered the situation of uniaxially, uniform, tensile strain described by a diagonal, coordinate independent strain tensor in the corresponding tight-bindig description of graphene. From the resulting dispersion relation, in the limit when the variation of the hopping energy due to the displacement of carbon atoms vanishes, such that the said dispersion relation is described in Eq.~(\ref{dispr}), within a semiclassical approximation, we simulate BO for different electric field orientations and varying the Poisson ratio, keeping the rest of the parameters of the model fixed. Feeding the ANN with 700 training signals, we were able to classify the components of the strain tensor for 300 new randomly generated signals. \\
For the case 1, when the component $E_y$ of the external electric field applied to the graphene is zero, the network has an accuracy in their predictions for $\epsilon_{xx}$ that goes from $79.0 \%$ to $88.6 \%$ with an error of $\pm 0.25\%$ and $\pm 1\%$ respectively.\\
Analogously, when $E_x=0$ the network has an accuracy in their predictions for $\epsilon_{xx}$ that goes from $68.3 \%$ to $90.3 \%$ with an error of $\pm 0.25\%$ and $\pm 1\%$ respectively.\\
When the external electric field has a random orientation, the accuracy predicting $\epsilon_{xx}$ goes from $54.6 \%$ to $81.6\%$ with an error of $\pm 0.25\%$ and $\pm 1\%$ respectively.\\
In the case 2, the network predict simultaneously $\epsilon_{xx}$ and $\nu$ for a random electric field orientation, obtaining an accuracy of $96.3\%$ with an error of $\pm10\%$ predicting $\epsilon_{xx}$ and an accuracy of $97.3\%$ with an error of $\pm25\%$ predicting $\nu$. This situation has the higher accuracy but also the higher error associated to each parameter. For the scenario with the lower error associated, the network has an accuracy of $84.3\%$ with an error of $\pm5\%$ predicting $\epsilon_{xx}$ and an accuracy of $87.0\%$ with an error of $\pm12.5\%$ predicting $\nu$. These results can be improved increasing the number of simulations that feed the ANN, but also increasing the computational time used during the network's training.\\

These encouraging results motivate us to pursue a more complete study of the full dispersion relation~(\ref{fulldr})~\cite{saul}. This is work in progress and results will be presented elsewhere.

\section{Acknowledgments}
The authors would like to thank Saul Hern\'andez-Ortiz for very useful discussions.
We acknowledge support from Consejo Nacional de Ciencia y Tecnolog\'ia (M\'exico) under grant 256494 and CIC-UMSNH under grant 4.23. We also thank for providing computer resources to ABACUS Laboratorio de Matem\'aticas Aplicadas y C\'omputo de Alto Rendimiento del CINVESTAV-IPN under grant CONACT-EDOMEX-2011-C01-165873.

 
\end{document}